\documentclass[preprint]{revtex4-1}
\usepackage{amsmath}
\usepackage{amscd}
\usepackage{amsfonts}
\usepackage{graphicx}
\usepackage{slashed}
\newcommand{\field}[1]{\mathbb{#1}}
\newcommand{\C}{\field{C}}
\newcommand{\R}{\field{R}}
\newcommand{\Z}{\field{Z}}
\newcommand{\Q}{\field{Q}}

\numberwithin{equation}{section}

\begin{document}

\title{The K-Theoretic Formulation of D-Brane Aharonov-Bohm Phases}

\author{Aaron R.~Warren}
\email{arwarren@purdue.edu}
\affiliation{Department~of~Mathematics,~Physics,~\&~Statistics\\
Purdue~University~North~Central,\\
1401~S.~US-421,~Westville,~IN~~46391}

\date{\today}

\begin{abstract}
The topological calculation of Aharonov-Bohm phases associated with D-branes in the absence of a Neveu-Schwarz B-field is explored.  The K-theoretic classification of Ramond-Ramond fields in Type II and Type I theories is used to produce formulae for the Aharonov-Bohm phase associated with a torsion flux.  A topological construction shows that K-theoretic pairings to calculate such phases exist and are well-defined.  An analytic perspective is then taken, obtaining a means for determining Aharonov-Bohm phases by way of the reduced eta-invariant.  This perspective is used to calculate the phase for an experiment involving the $(-1)-8$ system in Type I theory, and compared with previous calculations performed using different methods.
\end{abstract}

\maketitle

\section{Introduction} It is well known that the existence of a magnetic field will affect the phase of electrically-charged particles, even when the particles do not pass through the region containing the magnetic field.  The canonical example was formulated by Aharonov and Bohm \cite{abphase}, and is shown in Figure 1.  

\begin{figure}[h!]
	\centering
		\includegraphics[width=0.5\textwidth]{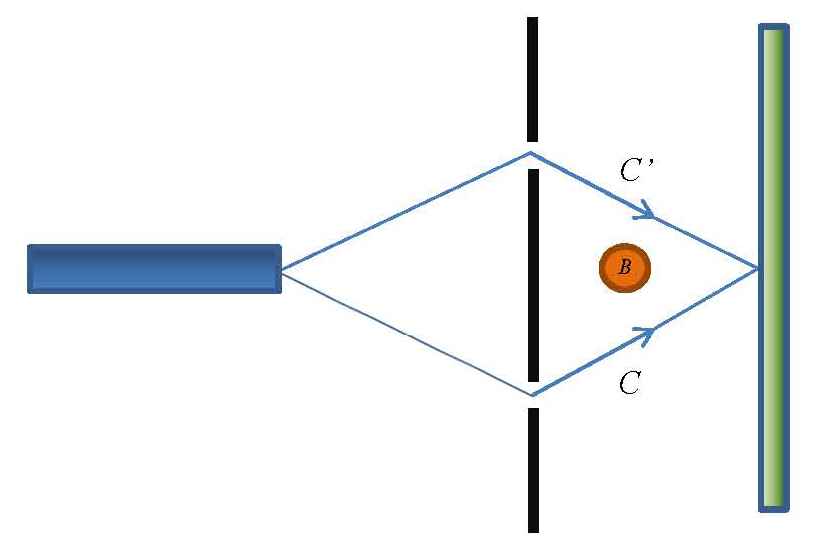}
	\caption{The canonical Aharonov-Bohm setup, with electrically-charged particles moving along each of two paths, \textit{C} and \textit{C'}, before interferring on a screen.}
\end{figure}

First, note that the existence of the $B$-field induces a connection $\omega = \frac{ie}{\hbar} A$, where $A$ is the Lie-algebra valued one form determined by the field.  Next, we let $\gamma = C -C' \in H_1 (X)$.  Then it is found that the phase acquired by a particle traveling along $\gamma$ is
\begin{equation}
\Phi[A,\gamma] = \frac{ie}{\hbar} \oint_\gamma A.
\end{equation}
Imposing the reality condition $\Phi[A_1,\gamma] \sim \Phi[A_2,\gamma]$ if $A_1-A_2 \in \Z$, the set of equivalence classes $[\Phi[A,\gamma]]$ are $\R/\Z \cong U(1)$-valued.  So we may view $[\Phi[A,\cdot]$ as an element of $H^1 (X;U(1))$.  Then we see that the Aharonov-Bohm phase is given by a pairing $H_1 (X) \times H^1(X;U(1)) \to U(1)$ defined as $[\Phi[A,\gamma]]$.

When we consider D-branes, however, things are not so simple.  In \cite{wit} and \cite{mw} it was shown that D-brane charges and Ramond-Ramond (RR) fields in Types IIA, IIB, and I theories are classified by K-theory.  Therefore, the calculation of Aharonov-Bohm phases for D-branes will necessarily involve some sort of K-theoretic pairing.

In this paper, Sections 2.1-2.4 produce a number of details concerning the topological formulation of D-brane Aharonov-Bohm phases in Type IIA theory, building off of a brief speculative discussion in \cite{mms}.  It is shown that the pairing outlined in \cite{mms} exists and is well-defined.  Section 2.5 provides adaptations of this pairing to the Type IIB and Type I settings.  In Section 3, the focus shifts to the use of the reduced eta-invariant as a means for calculating the K-theoretic pairing.  A brief overview of relevant mathematical technology is presented in Section 3.1 and then utilized in Section 3.2 for the $(-1)-8$ system in Type I theory.  It is shown that our result agrees with a calculation performed in \cite{guk} using different methods.   

\section{The Topological Formulation}

Let us begin by considering Type IIA theory on $\R_t \times X_9$, and suppose we have a brane producing a torsion flux.  This flux defines an element of $K^0_{tors} (X)$, the torsion subgroup of $K^0 (X)$, where $X \equiv X_8 = \partial X_9$.  

\subsection{The Long Exact Sequence}

We next wish to lift our element of $K^0_{tors} (X)$ to an element of $K^{-1} (X;U(1))$.  Before doing this, it is useful to consider the analogous cohomological situation.  From the exact coefficient sequence
\begin{equation}
0 \to \Z \to \R \to U(1),
\end{equation}
we may obtain the long exact cohomological sequence
\begin{equation}
\cdots \xrightarrow{\delta^{k-1}} H^k (X;\Z) \xrightarrow{i^k} H^k(X;\R) \xrightarrow{j^k} H^k (X;U(1)) \xrightarrow{\delta^k} H^{k+1} (X;\Z) \xrightarrow{i^{k+1}} H^{k+1} (X;\R) \to \cdots
\end{equation}
where the maps $\delta^k : H^k(X;U(1)) \to H^{k+1}(X;\Z)$ are called the Bockstein homomorphisms.

Since $\R/n\R \cong 0$ for any $n \in \Z-\{0\}$, the kernel of $i^{k+1}$ is the set of torsion elements of $H^{k+1} (X;\Z)$, denoted $H^{k+1}_{tors}(X;\Z)$.  Therefore, we may write the following exact sequence,
\begin{equation}
H^k (X;U(1)) \xrightarrow{\delta^k} H^{k+1}_{tors}(X;\Z) \to 0.
\end{equation}
Thus for any torsion class there is a lift in $H^k (X;U(1))$.  This lift is an integral cochain that is closed in $U(1)$ but not in $\Z$.  Simple diagram-chasing shows that this lift is well-defined \cite{mun}.

A completely analogous argument goes through in K-theory.  Indeed, one may write a long exact sequence similar to that above, and all subsequent statements and actions carry over.  Specifically, the long exact sequence
\begin{equation}
\cdots \to K^{-1}(X) \xrightarrow{ch} K^{-1}(X;\R) \xrightarrow{\alpha} K^{-1}(X;U(1)) \xrightarrow{\beta} K^0(X) \xrightarrow{ch} K^0(X;\R) \to \cdots
\end{equation}
gives the exact sequence
\begin{equation}
K^{-1}(X;U(1)) \xrightarrow{\beta} K^0_{tors}(X) \xrightarrow{ch} 0.
\end{equation}
Here, $ch$ is the Chern character and $\beta$ is the forgetful map.  Lifting an element of $K^0_{tors}(X)$ via the Bockstein $\beta$ then gives an element of $K^{-1}(X;U(1))$.  As in the cohomological case, diagram-chasing shows that this lift is well-defined.  

\subsection{The K-Cup Product}

Next, note that the test brane defines an element of $K^0(X)$.  We will pair this with our element of $K^{-1}(X;U(1))$ via the K-cup product \cite{hat,hus,spgeo}.  Again we begin by considering the analogous cohomological case.  There, one starts with a mapping
\begin{equation}
S^p(X) \times S^q(X;G) \xrightarrow{\cup} S^{p+q}(X;G)
\end{equation}
which assigns to each $p$-cochain $c^p$ and $q$-cochain $c^q$ a $(p+q)$-cochain $c^{p+q}$ by letting $c^p$ act on the front $p$-face and $c^q$ act on the $q$-back face, then multiplying the results by the usual product operation sending $(n,g)$ to $ng$.  It follows that $\cup$ gives a well-defined product operation \cite{mun}
\begin{equation}
H^p(X) \times H^q(X;G) \xrightarrow{\cup} H^{p+q}(X;G)
\end{equation}

In K-theory, we define a similar product operation via tensor products of bundles.  The external K-cup product is a group morphism $K(X) \otimes K(Y) \to K(X\times Y)$ which assigns to each $a \otimes b \in K(X)\otimes K(Y)$ the element $(K(p_x)(a))(K(p_y)(b)) \in K(X\times Y)$, where $p_x: X \times Y \to X$ and $p_y : X\times Y \to Y$ are the projection operators.  

When extended to higher K-groups, this product becomes $K^{-i}(X) \times K^{-j}(Y) \to K^{-i-j}(X \times Y)$.  To see this, we start with a pairing
\begin{equation}
\widetilde{K}^{-i}(X) \times \widetilde{K}^{-j}(Y) \to \widetilde{K}^{-i-j}(X \wedge Y)
\end{equation}
given by tensor product.  Next we use the defined relationship $K^{-i}(X) \equiv \widetilde{K}^{-i}(X^+) \equiv \widetilde{K}(\Sigma^i(X^+))$, where $X^+ \equiv X \cup \{pt.\}$ is $X$ with a disjoint basepoint and $\Sigma^i(X) = S^i \wedge X$ is the smash product of $S^i$ with $X$.  Then we obtain a pairing $K^{-i}(X)\times K^{-j}(Y) \to K^{-i-j}(X\times Y)$.

Letting $X=Y$ and then composing with the map from $K^{-i} (X \times X)$ to $K^{-i}(X)$ induced by the diagonal map $X \to X \times X$ gives the product 
\begin{equation}
K^{-i}(X) \times K^{-j}(X) \to K^{-i-j}(X).
\end{equation}
It follows that there is a pairing
\begin{equation}
K^{-i}(X) \times K^{-j}(X;G) \to K^{-i-j}(X;G),
\end{equation}
with $(n,g) \to ng$ as in the cohomological case.

We see that we may therefore pair 
\begin{equation}
K^0(X) \times K^{-1}(X;U(1)) \to K^{-1}(X;U(1))
\end{equation}
via the K-cup product with generalized coefficients.  Thus, given a brane producing a torsion flux and a charged test brane, we obtain an element of $K^{-1}(X;U(1))$.  

\subsection{K-Homology}

To measure the Aharonov-Bohm phase at infinity, we must move the test brane on a closed path in $X$.  This path defines an element of $H_1 (X)$.  In order to pair our path with $K^{-1}(X;U(1))$, we must lift the path to an element of $K_1(X)$, the K-homology of $X$ \cite{bm1}.

We may parametrize our path by a function $f: S^1 \to X$.  Note that $S^1$ is a compact $Spin^{\C}$-manifold without boundary, and $f$ is by definition a continuous map.  To put a complex vector bundle on $S^1$ is easy, since every complex vector bundle on $S^1$ is trivial.  It is natural, then, to let a K-cycle associated with our path be given by $(S^1, \epsilon^n, f)$ where $\epsilon^n$ is the trivial complex vector bundle with fibre $\C^n$.  Since $S^1$ is odd-dimensional, we have defined an element of $K_1(X)$.

To show that this lift is unique, we use the bordism and direct sum relations.  Consider the K-cycle $(S^1, \epsilon^2, f_2)$.  Since $\C^2 = \C \oplus \C$ the direct sum relation gives $(S^1, \epsilon^2, f_2) \sim (S^1, \epsilon^1, f_2) \cup (S^1, \epsilon^1, f_2)$.  Let $W$ be the compact 2-dimensional $Spin^\C$-manifold shown in Figure 2, and put the trivial rank 1 complex vector bundle on it.  

\begin{figure}[h!]
	\centering
		\includegraphics[width=0.5\textwidth]{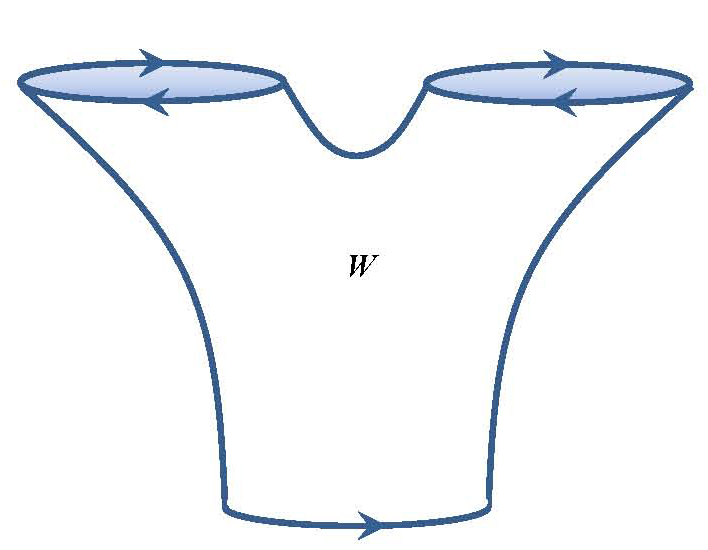}
	\caption{Surface $W$ carries the trivial rank 1 complex vector bundle on it, and serves as a bordism between $(-S^1, \epsilon^1, f_1)$ and two copies of $(S^1, \epsilon^1, f_2)$.}
\end{figure}

Then for appropriate choice of continuous $\phi : W \to X$, we have
\begin{equation}
(\partial W, E|_{\partial W}, \phi|_{\partial W}) \cong (S^1, \epsilon^1, f_2) \cup (S^1, \epsilon^1, f_2) \cup (-S^1, \epsilon^1, f_1) \sim (S^1, \epsilon^2, f_2) \cup (-S^1, \epsilon^1, f_1).
\end{equation}
Hence
\begin{equation}
(S^1,\epsilon^2,f_2) \sim (S^1, \epsilon^1,f_1).
\end{equation}
Clearly this generalizes to any K-cycle $(S^1, \epsilon^N,f_N)$.  Thus our lift from $H_1(X)$ to $K_1(X)$ is unique.

\subsection{The Intersection Form}

In the cohomological case, there is an intersection pairing on a compact oriented $n$-dimensional manifold $X$
\begin{equation}
H^k (X;\Z) \times H^{n-k}(X;U(1)) \to U(1)
\end{equation}
defined by
\begin{equation}
\alpha \times \beta \mapsto \alpha \cdot \beta \equiv \langle \alpha \cup \beta, [X]\rangle
\end{equation}
i.e., cup product followed by integration over an orientation class $[X] \in H_n (X;\Z)$.

We use this to define another pairing
\begin{equation}
H^k(X;\Z) \times H^{n-k+1}_{tors} (X;\Z) \to U(1)
\end{equation}
in the following way.  Since any torsion class $[\alpha] \in H^{n-k+1}_{tors}(X;\Z)$ has a well-defined lift $[\alpha']\in H^{n-k}(X;U(1))$, we may define the desired pairing as
\begin{equation}
\langle \beta \cup \alpha',[X]\rangle \in U(1).
\end{equation}

Returning to the K-theoretic case, recall that from our torsion flux, test brane, and path of the test brane we have defined elements of $K^1 (X;U(1))$ and $K_1 (X)$.  We would now like to pair these elements and get an element of $U(1)$, the Aharonov-Bohm phase.  This is achieved with the use of the so-called intersection form \cite{sav}.

The intersection form is the nondegenerate pairing
\begin{equation}
K^i_{cpt} (T^*X) \times K^i (X; U(1)) \to K^0_{cpt} (T^*X; U(1)) \xrightarrow{p_!} U(1)
\end{equation}
(with $cpt$ denoting compact support) which is induced by the K-cup product and the direct image mapping $p_! : K^0 (T^*X,U(1)) \to K^0 (pt,U(1)) = U(1)$ corresponding to the map $p: X \to pt$.

Poincar\'e duality and the Thom isomorphism give \cite{bm1,bm2},
\begin{equation}
K^1_{cpt} (T^*X) \cong K_1 (X).
\end{equation}
To see this isomorphism topologically, first let $S(T^*X)$ denote the unit sphere bundle of $T^*X$.  Also let $\pi:S(T^*X) \to X$ be the projection.  It was shown in \cite{aps3} that elements of $K^1(T^*X)$ are in one-to-one correspondence with stable homotopy classes of self-adjoint symbols on $X$.  Then an element of $K^1(T^*X)$ is a pair $(E,\sigma)$ where $E$ is a Hermitian vector bundle on $X$ and $\sigma:\pi^*(E) \to \pi^*(E)$ is a self-adjoint automorphism of $\pi^*(E)$.  Then $\sigma$ gives the decomposition $\pi^*(E) = E_+ \oplus E_-$ where $E_{\pm}$ is spanned by the eigenvectors with $\pm$ eigenvalues of $\sigma$.  

Setting $\widehat{X} = S(T^*X)$, note that $dim(\widehat{X}) = 2(dim(X)) - 1$ so that it is odd-dimensional.  Furthermore, $\widehat{X}$ is a $Spin^\C$-manifold since $T\widehat{X}$ has a $Spin^\C$ structure from $TX \oplus T^*X \cong \C \otimes_\R TX$.  Then the triple $(\widehat{X},E_+,\pi)$ is an element of $K_1(X)$.  If we define $c(E,\sigma) = (\widehat{X},E_+,\pi)$, then 
\begin{equation}
c: K^1(T^*X) \to K_1(X)
\end{equation}
is an isomorphism.

Thus the intersection form is a nondegenerate pairing between K-homology and K-theory,
\begin{equation}
K_1 (X) \times K^1 (X; U(1)) \to U(1).
\end{equation}
This is precisely what we need to give the Aharonov-Bohm phase.  

To summarize our formulation for the Type IIA case, the torsion flux defined an element of $K^0_{tors}(X)$ which we lifted to $K^{-1}(X;U(1))$ by the long exact sequence of K-groups associated with the exact coefficient sequence.  The test brane defined an element of $K^0(X)$ which we paired with our element of $K^{-1}(X;U(1))$ via the K-cup product to again get an element of $K^{-1}(X;U(1))$.  The path of our test brane defined an element of $H_1(X;\Z)$ which we lifted to an element of $K_1(X;\Z)$.  The intersection form then took $K_1(X;\Z) \times K^1(X;U(1)) \to U(1)$, which we call the Aharonov-Bohm phase.

\subsection{The Type IIB and Type I Cases}

Now that we have given the topological details of the K-theoretic formula for Aharonov-Bohm phase in the Type IIA case, we would like to develop similar statements for the Type IIB and Type I cases.

In the Type IIB situation, the torsion flux takes values in $K^1_{tors}(X)$, with $X = \partial X_9$ as before.  Then we may again use the exact coefficient sequence to give a long exact sequence of K-groups and lift our element of $K^1_{tors} (X)$ to $K^0 (X; U(1))$.  Now our test brane defines an element of $K^1(X)$, and we again use the K-cup product to pair these elements as 
\begin{equation}
K^1(X) \times K^0 (X;U(1)) \to K^1 (X;U(1)).
\end{equation}
Again we lift the path of the test brane from $H_1(X)$ to $K_1(X)$ and use the intersection form to pair
\begin{equation}
K_1(X) \times K^1(X;U(1)) \to U(1).
\end{equation}

Finally, we may make a similar proposal in the Type I scenario.  Here, the torsion flux is valued in $KO^{-1}_{tors}(X)$, and the test brane defines an element of $KO^{-1}(X)$.  All the properties of the complex K-theory that we employed carry over to the KO-groups.  The only real difference between these theories is the form of Bott Periodicity, but that does not seriously affect our discussion.  So we lift the torsion flux to $KO(X;U(1))$, then pair the test brane charge to it 
\begin{equation}
KO^{-1}(X) \times KO(X;U(1)) \to KO^{-1}(X;U(1))
\end{equation}
by a KO-cup product which is completely analogous to the K-cup product.  We can complexify to obtain an element of $K^{-1}(X;U(1))$.  Now we lift the path of the test brane from $H_1(X)$ to $K_1(X)$.  Then we use the intersection form to pair
\begin{equation}
K_1(X) \times K^{-1} (X;U(1)) \to U(1),
\end{equation}
giving us the Aharonov-Bohm phase.

Some explanation is required to justify labelling this $U(1)$ as an Aharonov-Bohm phase.  The question is whether this phase occurs within the partition function for a D-brane that participates in an Aharonov-Bohm experiment.  The interaction between the pair of D-branes involved in such an experiment will be mediated by open strings connecting the two branes, and to produce an Aharonov-Bohm phase they must be sensitive to their relative orientations.  Only fermions which become massless when the branes coincide are capable of detecting their relative orientations.  While the Neveu-Schwarz sector open string zero point energy is sometimes greater than zero \cite{pol}, Ramond-sector open strings always have a zero point energy equal to zero.  Therefore, there will be massless fermions whose sensitivity to relative orientation will affect the partition function, generating an Aharonov-Bohm phase.  

In the Type I case, these open string interactions can be viewed from the perspective of the effective gauge theory defined on the worldvolume of 9-branes used to construct the D-brane system.  The two D-branes correspond to topological defects in the gauge bundle defined on the 9-brane system, and the K-theoretic pairing specified above measures the topological phase induced by the relative motion of the defects.  We shall return to this gauge bundle perspective later, in Section 3.2.

\section{Analytical Aspects}
Here we describe the pairing
\begin{equation}
K_1(X;\Z) \times K^1(X;U(1)) \to U(1)
\end{equation}
from an analytic point of view.  We begin by reviewing relevant material from \cite{lot} to define the reduced eta-invariant, and relate it to the topological pairing from Section 2.  We then use the eta-invariant to calculate the phase for an Aharonov-Bohm experiment involving a $(-1)-8$ brane system in Type I theory.  

\subsection{The Analytic Formulation}

First we define a $\Z_2$-graded cocycle in $K^1(X;U(1))$ to be a quadruple $\mathcal{V} = (V_{\pm},h^{V_{\pm}},\nabla^{E_{\pm}},\omega)$, where: 
\begin{itemize}
\item $V = V_+ \oplus V_-$ is a $\Z_2$-graded vector bundle on $X$ 
\item $h^V = h^{V_+} \oplus h^{V_-}$ is a Hermitian metric on $V$ 
\item $\nabla^V = \nabla^{V_+} \oplus \nabla^{V_-}$ is a Hermitian connection on $V$
\item $\omega \in \Omega^{odd}/im(d)$ satisfies $d\omega = ch_{\Q}(\nabla^V)$.
\end{itemize}
We may define a $Z_2$-graded cocycle in $KO^1(X;U(1))$ analogously, by replacing the adjectives ``complex'' and ``Hermitian'' with ``real'' and ``symmetric,'' respectively.

Next, recall that a K-cycle in $K_1(X)$ is a triple $(M,E,f)$ with $M$ a closed odd-dimensional $Spin^{\C}$-manifold, $E$ a complex vector bundle on $M$, and $f:M \to X$ a continuous map.  We will in fact let $f$ be smooth here.  Again note that there is an analogous real formulation.  For the rest of this subsection, however, we'll restrict our attention to the complex case.

Since $M$ is $Spin^{\C}$, the principle $GL(dim(M))$-bundle on $M$ may be reduced to a principle $Spin^{\C}$-bundle, call it $P$.  We may associate to $P$ a Hermitian line bundle $L$ on $M$ \cite{fri}.  Choose a Hermitian connection $\nabla^L$ on L, a Hermitian metric $h^E$ on $E$, and a Hermitian connection $\nabla^E$ on $E$.

Let $\widehat{A}(\nabla^{TM}) \in \Omega^{even}(M)$ be the closed form representing $\widehat{A}(TM) \in H^{even}(M;\Q)$.  Also, let $\exp[c_1(\nabla^L)/2] \in \Omega^{even}(M)$ be the closed form representing $\exp[c_1(L)/2] \in H^{even}(M;\Q)$.  Finally, let the spinor bundle of $M$ be $S_M$.

Then given a $\Z_2$-graded cocycle $\mathcal{V} \in K^1(X;U(1))$, we let $D_{f^* \nabla^{V_{\pm}}}$ denote the Dirac-type operator acting on sections of $S_M \otimes E \otimes f^*V_{\pm}$.  Its reduced eta-invariant is \cite{aps2}
\begin{equation}
\bar{\eta}(D_{f^* \nabla^{V_{\pm}}}) = \frac{1}{2}[\eta(D_{f^* \nabla^{V_{\pm}}}) + dim(Ker(D_{f^* \nabla^{V_{\pm}}}))]~~~ mod~~\Z.
\end{equation}
Then the reduced eta-invariant of $f^*\mathcal{V}$ is the $\R/\Z$-valued function
\begin{equation}
\bar{\eta}(f^*\mathcal{V}) = \bar{\eta}(D_{f^* \nabla^{V_+}}) - \bar{\eta}(D_{f^* \nabla^{V_-}}) - \int_M \widehat{A}(\nabla^{TM}) \wedge \exp[c_1(\nabla^L)/2] \wedge ch_{\Q} (\nabla^E) \wedge f^*\omega.
\end{equation}

Finally, given a cycle $\mathcal{K} = (M,E,f)$ in $K_1(X)$ and a $\Z_2$-graded cocycle $\mathcal{V}$ for $K^1(X;U(1))$, their $\R/\Z$-valued pairing is
\begin{equation}
\langle [\mathcal{K}],[\mathcal{V}] \rangle = \bar{\eta}(f^*\mathcal{V}).
\end{equation}
The proof that this is in fact the correct pairing is given in \cite{lot}, and is based largely on the corresponding proof in \cite{aps2}.  We claim that this yields the Aharonov-Bohm phase as
\begin{equation}\label{pairing}
2\pi i ~ \bar{\eta}(f^* \mathcal{V}).
\end{equation}

Note that we can use the D-brane charges instead of the RR fields in our prescriptions.  The K-theory class associated with an RR-field may be mapped in a well-defined way to the K-theory class associated with the D-brane charge via the isomorphism
\begin{equation}
K^i_{cpt}(M) \cong K^{i-1}(\partial M)/j(K^{i-1}(M)),
\end{equation}
where $j$ restricts a K-theory class from $M$ to $\partial M$.  The analogous isomorphism holds for the KO-theory as well.  Therefore, if desired we may change our prescriptions to begin with the K-theory classes associated with the charges of the D-branes.  

\subsection{Calculation in the Type I Case}

We now consider an Aharonov-Bohm experiment for a $(-1)-8$ system of Type I D-branes.  The path of the instanton defines the K-cycle $(M,\epsilon^1,f)$ as discussed above.  We use the 32 nine-branes required for tadpole cancellation to construct our system without adding extra branes/antibranes.  It will be convenient to work from the K-theory classes of the D-brane charges instead of those of the RR fields.

The 8-brane determines the non-trivial element of $KO^0_{tors}(S^1)=\Z_2$, which we view as $KO^0_{tors}(\R^1)$ with compact support.  Such an element is given by the pair $(E_8, pt)$ where $E_8$ is a rank 1 bundle and $pt$ is the trivial rank 0 bundle.  We lift this via the Bockstein to an element $(E'_8,pt) \in KO^{-1}(\R;U(1))$.  Note that $E'_8$ is also a rank 1 bundle.

Next, the (-1)-brane determines an element of $KO^0(S^{10}) = \Z_2$.  This is the pair $(E_{-1},pt)$.  Taking the KO-cup product we get $(E_{-1} \otimes E'_8,pt) \in KO^{-1}(\R^{11};U(1))$.  After complexifying these bundles, we then obtain a $\Z_2$-graded cocycle $V = (E_{-1} \otimes E'_8 \otimes \C) \oplus pt$ in $K^{-1}(\R^{11};U(1))$.

We also have the associated Dirac operator along $M$ for the $E_{-1} \otimes E'_8 \otimes \C$ component of the cocycle,
\begin{equation}
\slashed{D}^+_a = \slashed{D}(A'_{-1}) + \slashed{D}(A_8) + \Gamma^9 a
\end{equation}
where $a$ parametrizes $M$ as the distance between the 8-brane and the (-1)-brane as above, and the $+$ superscript indicates that it corresponds to the $V_+ \equiv E_{-1} \otimes E'_8 \otimes \C$ component.  Since $E'_8$ is a rank 1 bundle, the index theorem \cite{aps1, aps2} says that $\slashed{D}(A'_{-1}) + \slashed{D}(A_8)$ has one zero mode of definite chirality with respect to $\Pi_{\mu=0}^9 \Gamma^\mu$ and $\Gamma^9$ (in the non-trivial instanton number sector).  Thus $\slashed{D}^+_a$ has eigenvalue $a$.

Our analytic pairing can be evaluated by using the pullback via $f$ of $V =(E_{-1}\otimes E'_8 \otimes \C) \oplus pt$ from $X$ to $M$.  Since the eigenvalue of $\slashed{D}^+_a$ is equal to $a$, we get $\bar{\eta}(D_{f^* \nabla^{V_+}}) = \frac{1}{2}$.  Also, since $E_{-1}$ and $E'_8$ are each $SO(n)$-bundles, their connection and curvature forms are $so(n)$-valued, hence have zero trace.  Complexification does not change this, so that the Chern character of $V$ is zero.  Then since $d\omega = ch_\Q(\nabla^V) = 0$ and $\omega \in \Omega^{odd}/Im(d)$, we get $\omega = 0$ and $f^*\omega = 0$.  Finally, note that $\bar{\eta}(D_{f^*\nabla^{V_-}}) = 0$ since $V_- = pt$.

Then we find that our pairing gives
\begin{equation}
\bar{\eta}(f^*\mathcal{V}) = \bar{\eta}(D_{f^*\nabla^{V_+}}) - \bar{\eta}(D_{f^*\nabla^{V_-}}) - \int_{M} \widehat{A}(\nabla^{TM}) \wedge e^{\frac{c_1(L)}{2}} \wedge ch_\Q (\nabla^{\epsilon^1}) \wedge f^* \omega = \frac{1}{2}.
\end{equation}
Consequently, we get an Aharonov-Bohm phase of $2 \pi i \frac{1}{2} = i\pi$, and the monodromy is $\exp(i\pi) = -1$.  

This result agrees with the calculation performed in \cite{guk} for the $(-1)-8$ system, which was performed by examining changes in massless fermionic contributions to the amplitude as the instanton is moved.

\section{Summary}

In this paper we have developed formulae to calculate the Aharonov-Bohm phase of torsion Ramond-Ramond fluxes in the Type II and Type I string theories based upon the K-theoretic classification of Ramond-Ramond fields and D-brane charges.  These formulae were constructed in two different but equivalent fashions, one being purely topological and the other employing the reduced eta-invariant.  The topological pairing was shown to exist and be well-defined.  The analytic perspective was used to calculate the phase for the $(-1)-8$ system in Type I theory, allowing us to test our forumlae by comparison with independent calculations.

\begin{acknowledgments}
The author would like to thank G.W. Moore for helpful discussions.  
\end{acknowledgments}

\end{document}